\documentclass[10pt]{iopart}

\usepackage{epsfig,iopams}

\begin{document}

\title[Riishede, Mortensen \& L\ae gsgaard, A ``poor man's approach'' to
modelling of ....]{ A ``poor man's approach'' to modelling of
micro-structured optical fibres}

\author{Jesper Riishede} 

\address{Research Center COM, Technical University of Denmark, DK-2800
Kgs. Lyngby, Denmark}

\author{Niels Asger Mortensen} 

\address{Crystal Fibre A/S, Blokken 84, DK-3460 Birker\o d, Denmark}

\author{Jesper L\ae gsgaard} 

\address{Research Center COM, Technical University of Denmark, DK-2800
Kgs. Lyngby, Denmark}

\begin{abstract}
Based on the scalar Helmholtz equation and the finite-difference
approximation, we formulate a matrix eigenvalue problem for the
calculation of propagation constants, $\beta(\omega)$, in
micro-structured optical fibres.  The method is applied to
index-guiding fibres as well as air-core photonic bandgap fibres, and
in both cases qualitatively correct results are found. The strength of
this approach lies in its very simple numerical implementation and its
ability to find eigenmodes at a specific eigenvalue, which is of great
interest, when modelling defect modes in photonic bandgap fibres.
\end{abstract}

\submitto{\JOA}

\section{Introduction}
Since the first results on photonic crystal fibres~\cite{knight1996}
(PCF) many exciting phenomena have been reported (for a recent review
see {\it e.g.} Ref.~\cite{russell2003} and references therein). From
the very start, a great emphasis has been on the modelling of the
optical properties and frameworks of high complexity have been
developed. In this work, we develop a ``poor man's approach'' which
allows for easy implementation and calculation of the propagation
constant, $\beta(\omega)$, in arbitrarily microstructured fibres.
Most approaches start from the fundamental vectorial wave equation for
an isotropic dielectric medium

\begin{equation}
{\boldsymbol \nabla}\times 
\frac{1}{\varepsilon({\boldsymbol r})}{\boldsymbol \nabla}\times 
{\boldsymbol H}({\boldsymbol r})=k^2{\boldsymbol H}({\boldsymbol r})
\end{equation}
where $k=\omega/c$ and $\varepsilon({\boldsymbol r})$ is the
dielectric function, which may be frequency dependent. For a fibre
geometry with $\varepsilon({\boldsymbol r})=\varepsilon(x,y)$, {\it
i.e.}, translational invariance along the $z$-direction, we look for
solutions of the form ${\boldsymbol H}({\boldsymbol r})={\boldsymbol
h}(x,y)e^{i\beta z}$. Substituting this ansatz into the wave equation
results in an eigenvalue problem, which determines the dispersion
$\omega(\beta)$. In the literature, it is often emphasized that in
general a fully-vectorial approach is required for quantitative
modelling of micro-structured fibres. Several fully-vectorial
approaches have been reported including plane-wave
methods~\cite{ferrando1999,johnson2001}, multi-pole
methods~\cite{white2002,kuhlmey2002}, localised-function
methods~\cite{mogilevtsev1999,monro2000}, and finite-element
methods~\cite{koshiba2001}. The complicated implementation is common
to all these methods. In this work, we present a ``poor man's
approach'' for calculating the propagation constant, $\beta(\omega)$,
in arbitrary dielectric structures. Despite its obvious shortcomings, we
believe it is far more useful for more qualitative studies as well as
for teaching of the physics of micro-structured optical fibres at the
introductory level.

\section{Theory}

\subsection{The scalar Helmholtz wave-equation}

We start from the scalar Helmholtz wave-equation 

\begin{equation}\label{Helmholtz}
\left(\frac{\partial^2}{\partial x^2}+\frac{\partial^2}{\partial y^2}+
\varepsilon(x,y)k^2\right) \Psi(x,y)=\beta^2 \Psi(x,y)
\end{equation}
with $\Psi$ being the scalar field. In this approximation, we have
neglected a logarithmic derivative of the dielectric function at the
interfaces between {\it e.g.} SiO$_2$ and air. We also note that
polarisation effects and/or degeneracies may only be fully revealed by
a fully-vectorial approach. The strength of the scalar approach is that it
is straight forward to implement and thus serves as an excellent
starting point for researchers and students entering the field of
micro-structured optical fibres. Furthermore, the Helmholtz equation
(\ref{Helmholtz}) allows for an easy incorporation of effects like
material dispersion, $\varepsilon(\omega)$.

\begin{figure}[t!]
\begin{center}
\epsfig{file=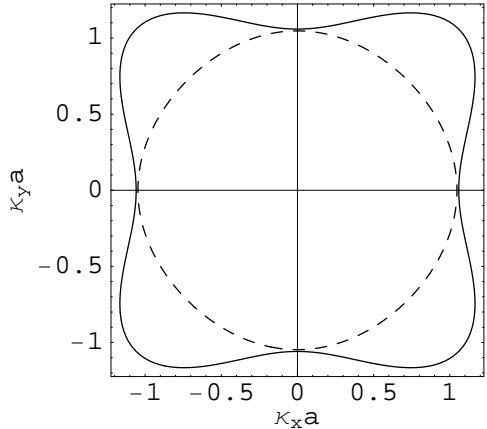, width=0.5\columnwidth,clip}
\end{center}
\caption{Contour plot of Eq.~(\ref{homogeneous}) for $\omega^2=
\frac{c^2}{\varepsilon}\left[\beta^2 + K^2\right]$ shown by the dashed
line. For the region enclosed by the solid line, the relative error of
the finite-difference approximation is generally less than $9\,\%$, and
the relative error is zero at the origo ($\kappa_x=\kappa_y=0$) where
$\omega^2= \frac{c^2}{\varepsilon}\beta^2$. }
\label{fig1}
\end{figure}

\subsection{The finite-difference approximation}

Equation (\ref{Helmholtz}) constitutes an eigenvalue problem from
which $\beta(\omega)$ can be calculated. We take the probably most
simple approach based on a finite difference approximation of the
Laplacian. For a quadratic grid with $j$ labeling the grid points with spacing $a$ we {\it e.g.} get the
symmetrized representation of the Laplacian

\begin{equation}
\frac{\partial^2}{\partial x^2} f[x=ja]\approx \frac{1}{a^2}\left(
f[(j+1)a]+ f[(j-1)a] -2f[ja] \right)
\label{findif-laplace}
\end{equation}
corresponding to nearest-neighbour coupling of the grid points. We can
thus rewrite Eq.~(\ref{Helmholtz}) as a matrix eigenvalue problem

\begin{equation}
{\boldsymbol \Theta}{\boldsymbol \Psi}=\beta^2{\boldsymbol \Psi}
\end{equation}
with
\begin{equation}
{\boldsymbol \Theta}_{ji}= \left\{ \begin{array}{cc} -4K^2
  +\varepsilon_j k^2 & j=i\\ K^2& j,i\, {\rm nearest\,neighbours}\\ 0 &
  {\rm otherwise}\end{array}\right.
\label{findif-theta}
\end{equation}
where $K=1/a$. For the dielectric function we have $\varepsilon_j=\varepsilon(x_j,y_j)$ where $(x_j,y_j)$ is the coordinates of the $j$th lattice point. The numerical task thus consists of finding eigenvalues
of the matrix $\boldsymbol \Theta$, which is easily done using standard
numerical libraries. The matrix is highly sparse, symmetric and when
the dielectric function is real, it is even Hermitian. The numerical
accuracy of the approximation is of course increased by decreasing the
lattice constant.

\subsection{The homogeneous case}
In order to estimate the required size of $a$, we first consider
the homogeneous case where $\varepsilon_j=\varepsilon$. This problem
is well-known from solid state theory; it can be diagonalized by a
plane wave ansatz, which results in the usual cosine-band result
 
\begin{equation}\label{homogeneous}
\omega^2 = \frac{c^2}{\varepsilon}\left[\beta^2 +
2K^2(2-\cos\kappa_xa-\cos\kappa_ya)\right].
\end{equation}
This result also has the correct asymptotic behaviour of the
homogeneous-space solution

\begin{equation}
\omega^2 \simeq\frac{c^2}{\varepsilon} \left[ \beta^2 +
\kappa_x^2+\kappa_y^2\right] + {\cal O}( a^2)
\end{equation}
and by choosing $a$ sufficiently small, the numerical discretisation is
a good approximation to the solution of the exact problem. Because of
the discretisation procedure, Eq.~(\ref{homogeneous}) has a finite
band-width of $\frac{c^2}{\varepsilon}\times 8K^2$, {\it i.e.},
\begin{eqnarray}
\max \{2K^2(2-\cos\kappa_xa-\cos\kappa_ya)\}\\ \;\;\;-\min
\{2K^2(2-\cos\kappa_xa-\cos\kappa_ya)\}=8K^2.\nonumber
\end{eqnarray}
This means that only frequencies satisfying 
\begin{equation}
\frac{c^2}{\varepsilon}\beta^2 \leq \omega^2\leq
\frac{c^2}{\varepsilon}\left[\beta^2 + 8K^2\right]
\end{equation}
can be accounted for by the discretisation procedure. In the
low-frequency regime, the relative error of the finite-difference
approximation becomes small, and typically we should be limiting
ourselves to {\it e.g.},
\begin{equation}
\omega^2< \frac{c^2}{\varepsilon}\left[\beta^2 +
K^2\right]\Leftrightarrow a <
\left(\varepsilon\omega^2/c^2-\beta^2\right)^{-1/2}
\end{equation}
where the relative error of the finite-difference approximation is
less than $9\,\%$, see Fig.~\ref{fig1}. For higher frequencies, the
finite-difference procedure becomes an inaccurate approximation to the
exact problem, because of artificial band-structure effects.

\subsection{The boundary problem}
In principle $\boldsymbol \Theta$ is infinite and for the
implementation we obviously need to truncate the matrix. This
truncation may affect the accuracy of the calculation. However, we are
also faced with the problem of deciding how the finite-difference
representation of the differential operators (in our case the
Laplacian) are represented on the boundary of the calculation domain.
This problem arises because calculation of finite differences on the
boundary requires the use of points that fall outside the calculation
domain. Thus, we have to determine a proper way of how these
non-existing points should be treated.

In the definition of the $\boldsymbol \Theta$-matrix, we have simply
chosen to neglect the points that fall outside the calculation domain.
This is done in order to limit the complexity of the $\boldsymbol
\Theta$-matrix, and thereby keep the formulation of the problem as
simple as possible. The consequence of this simplification is that we
impose the condition that the field has to be zero on the boundaries
of the calculation domain. Naturally, this assumption will have an
effect on the accuracy of the calculation, but the better the field is
confined within the calculation domain, the better the truncated
problem resembles the correct solution, since a zero amplitude on
the boundary becomes a reasonable approximation in this case.

\section{Modelling of Photonic Crystal Fibres}
\subsection{Numerical Implementation}
Once the theory of the finite difference approximation has been
established, the task of finding solutions to the scalar wave equation
may be viewed as two subproblems.  First, the $\boldsymbol
\Theta$-matrix has to be created from a given dielectric structure,
and secondly the eigenvalues, $\beta^2$, and the corresponding
eigenvectors, $\boldsymbol \Psi$, have to be found. We have chosen to make
our implementation in \emph{Matlab}, because it provides effective tools
for solving both these problems.

To give a more precise description of our implementation, we consider
an example where an arbitrary dielectric structure has been discretisized in
a 100 $\times$ 100 grid. In this case $\boldsymbol \Theta$ becomes a
matrix with 10000$\times$10000 elements, which indicates that the
finite difference approach is very demanding in terms of memory
consumption.  However, as most of the elements of $\boldsymbol \Theta$
are zero, it is advantageous to store $\boldsymbol \Theta$ as a sparse
matrix, which is easily done with the \emph{Sparse-type} in
\emph{Matlab}. For an n $\times$ n dielectric structure, we need to store
$n^4$ elements in the full representation, while only $\sim5n^2$
elements are required in the sparse representation. Obviously, this
gives rise to a dramatic decrease in the memory consumption, and
thereby a corresponding increase in the size of the dielectric
structures that may be examined.

The second problem we are faced with, concerns the search for
eigenvalues, $\beta^2$, and their corresponding eigenvectors
$\boldsymbol \Psi$. In the sparse representation, a complete
diagonalization of the $\boldsymbol \Theta$-matrix may be performed,
but this straightforward method has the disadvantage of being
numerically heavy (and thus time consuming) since it calculates all
$n^2$ eigenvalues. Typically, we are only interested in solving for a few
eigenvalues, e.g the largest values of $\beta^2$, and this is a common
feature of several advanced eigensolver libraries. In our
implementation we use the \emph{EIGS} command in \emph{Matlab} which is based
on the ARPACK-library \cite{arpack}. As a further advantage, the
\emph{EIGS} command has the option of finding eigenvalues around a
specified value, which may be particularly useful once a region with
guided modes has been found.

\subsection{Index-guiding Fibres}
\begin{figure}[htpb]
\begin{center}
\epsfig{file=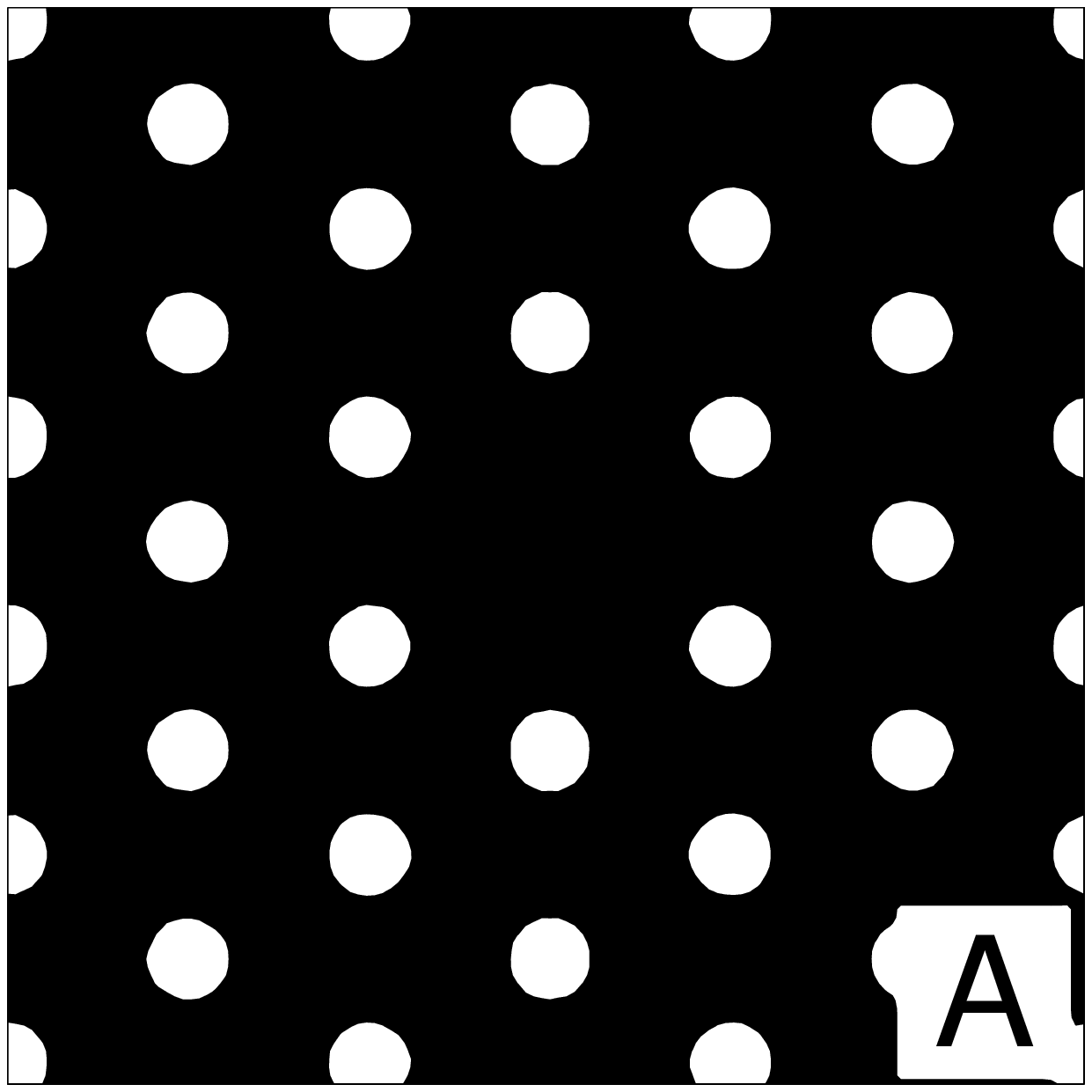, width=0.4\columnwidth,clip}
\hspace{0.4cm}
\epsfig{file=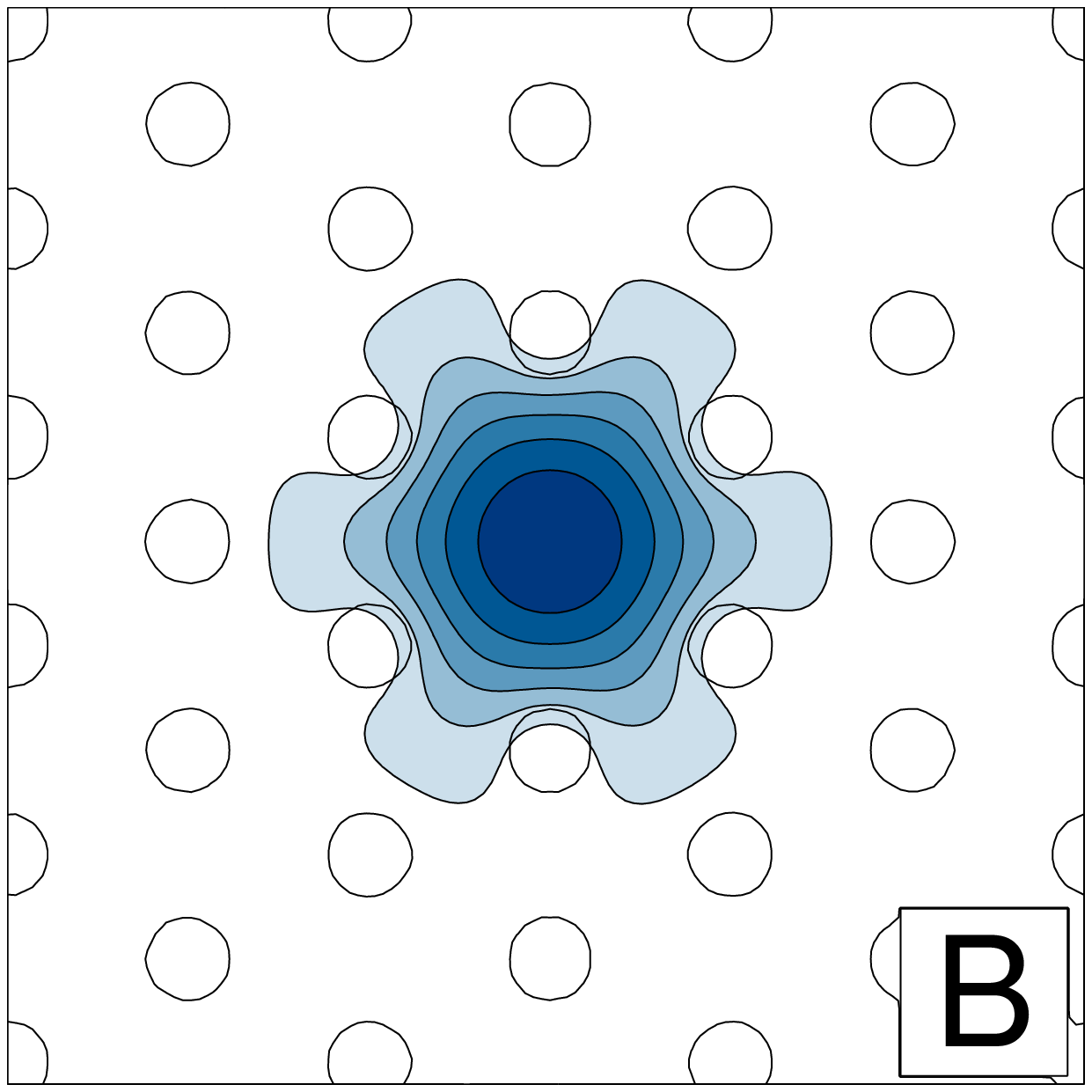, width=0.4\columnwidth,clip}
\end{center}
\caption{(A) A dielectric structure of an index-guiding photonic
crystal fibre, with a normalized holediameter of $D/\Lambda = 0.4$. For
the calculation the structure has been discretized in $128 \times 128$
points. (B) The field distribution of the fundamental mode, calculated
at a normalized wavelength of $\lambda/\Lambda = 0.15$.}
\label{pcfmode}
\end{figure}

As a first example, the finite-difference method is applied to an
index-guiding micro-structured fibre. Figure \ref{pcfmode}a shows the
square dielectric structure used in the calculation, which we have
chosen to discretisize in a 128 $\times$ 128 grid.  The dielectric
structure has a width of $3 \sqrt{3} \Lambda$, where $\Lambda$ is the
hole spacing, and it consists of air holes with a diameter of
0.4$\Lambda$ placed in a silica background (n=1.45). A single air hole
has been omitted in the center of the structure to create a waveguide
core. In the case of index-guiding fibres, the fundamental mode
corresponds to the eigenvector with the largest eigenvalue. Figure
\ref{pcfmode}a shows the field distribution of the fundamental mode
calculated at a normalized wavelength of $\lambda / \Lambda = 0.15$, and
it is seen to be well confined to the core region.

\begin{figure}[htpb]
\begin{center}
\epsfig{file=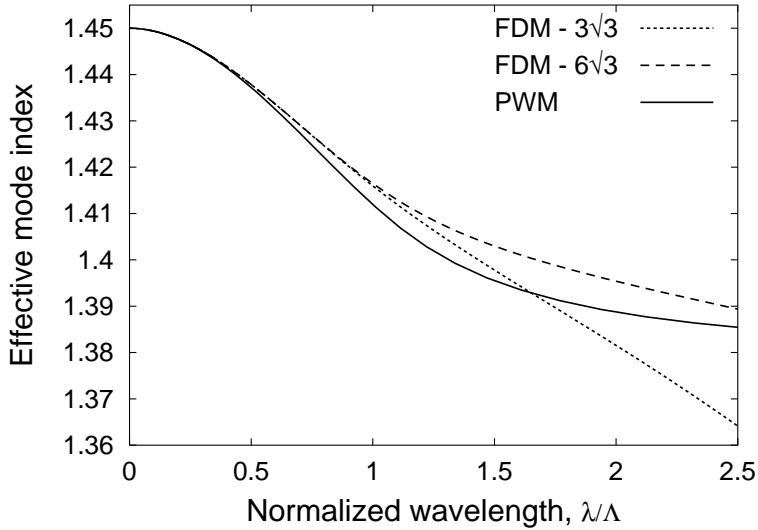, width=0.8\columnwidth,clip}
\end{center}
\caption{Comparison of the mode-index for the fundamental mode in an
index-guiding PCF with a hole diameter of $0.4 \Lambda$. The dotted
curves are calculated by the finite-difference method (FDM) for two
different widths of the dielectric structure, while the solid curve is
calculated by the plane-wave method (PWM).}
\label{pcfcompare}
\end{figure}

In figure \ref{pcfcompare} we have mapped out the effective mode index
of the fundamental mode for several values of the normalized
wavelength. For comparison, we have included a finite-difference
calculation, where the width of the calculation domain has been
increased to $6 \sqrt{3} \Lambda$. Finally, we have included
fully-vectorial results for a fully-periodic hole-structure, with a
repeated defect, obtained in a plane-wave basis using periodic boundary
conditions~\cite{johnson2001}. The dielectric structure used in this
calculation consist of 6 simple cells, and is thus similar to the
structure in figure \ref{pcfmode}a.

By comparing the finite-difference calculations for the small and the
large calculation domain, we are able to see the effect of the
truncation. For the small calculation domain, we find that the
calculated value of the effective mode-index decreases rapidly for
long normalized wavelengths. This is because the field distribution
extends to the edge of the calculation domain, and thus the assumption
of a zero amplitude on the edge is no longer valid. Consequently, the
field will penetrate into the air holes, and thereby cause a lowering
of the effective mode index. By increasing the width of the
calculation domain, we are able to shift the onset of this effect
towards larger values of the normalized wavelength.
 
In the comparison between our scalar finite-difference approach and a
full-vectorial method, we find that the scalar approach gives
qualitative correct results and accounts well for the overall
wavelength dependence. However, especially for the long wavelengths
the simple approach becomes inaccurate because of the scalar
approximation, and the scalar approach is seen to overestimate the
value of the effective mode-index. For $\lambda\sim \Lambda$ the
strong proximity of the air-holes require a correct treatment of the
air-silica boundary conditions which can only be quantitatively
accounted for by a fully-vectorial approach.

\subsection{Photonic Bandgap-guiding Fibres}
In the literature, it is often argued that accurate modelling of
photonic bandgap materials requires the use of a full-vectorial
method. This is true for all photonic crystals of practical interest,
because it is required that they have a large index-contrast in order to obtain
large bandgaps. However, this may lead to the incorrect conclusion
that photonic bandgaps and defect modes are pure full-vectorial
phenomena. From a theoretical point of view bandgaps do not arise as a
consequence of coupling between field components at a dielectric
interface, as it is the case in a full-vectorial method. Rather, they
are a fundamental property of applying the wave equation to a periodic
waveguide structure, and thus photonic bandgaps and defect modes can
also exist in a scalar calculation.

\begin{figure}[htbp]
\begin{center}
\epsfig{file=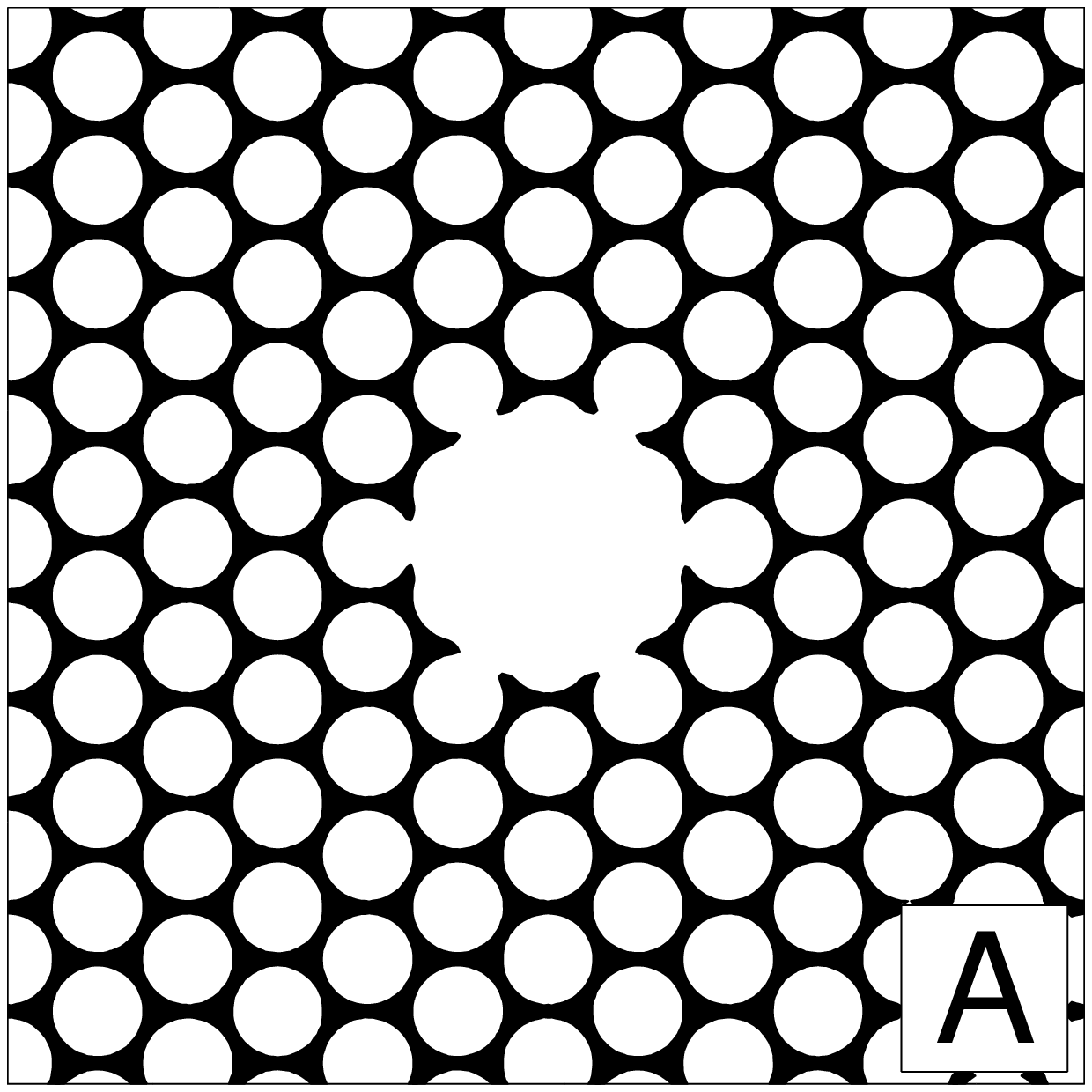, width=0.4\columnwidth,clip}
\hspace{0.4cm}
\epsfig{file=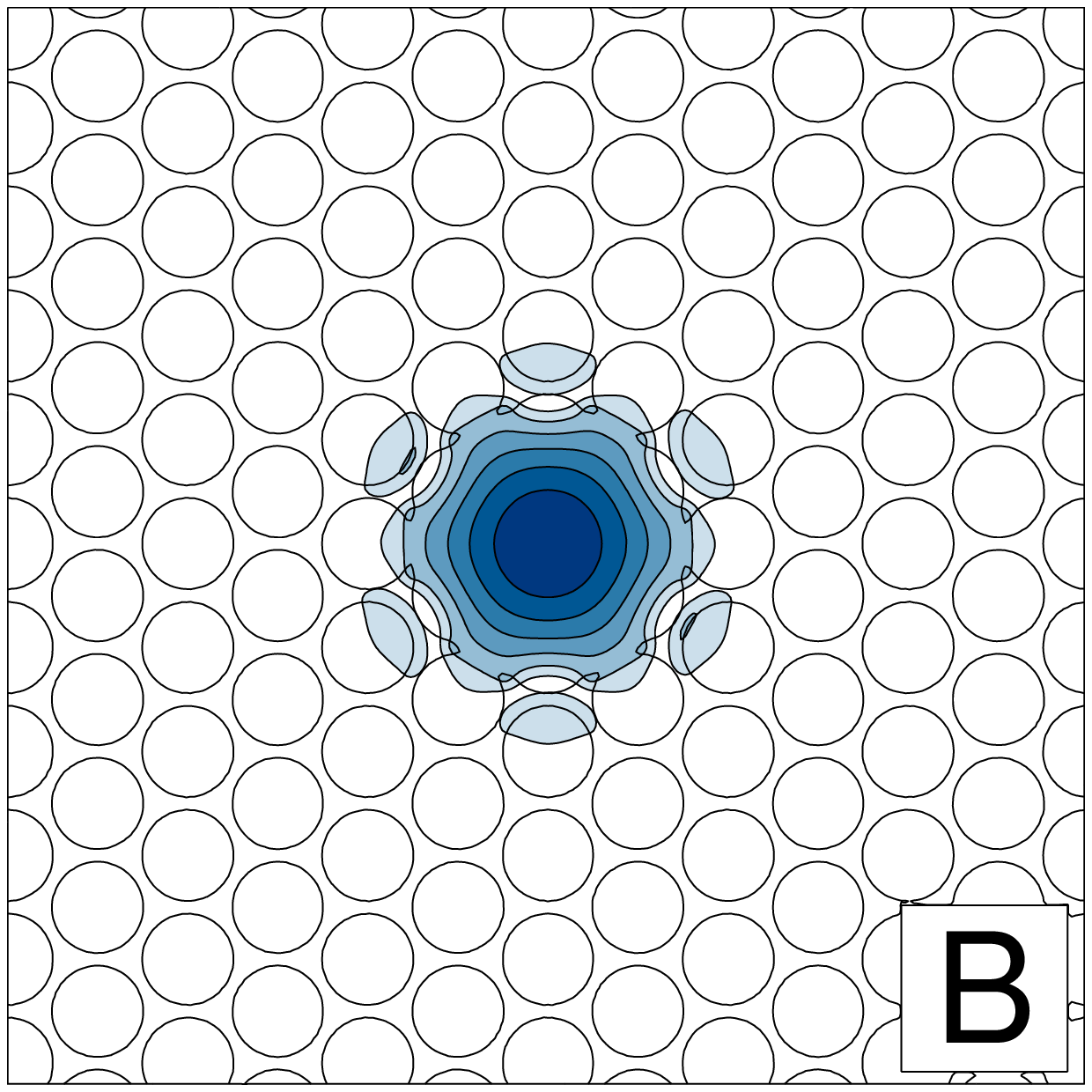, width=0.4\columnwidth,clip}
\end{center}
\caption{(A) The dielectric structure of an airguiding photonic
  bandgap fibre discretized in $256 \times 256$ points. The cladding
  structure has a hole diameter of $0.88 \Lambda$ while the core
  defect is created by inserting a large air hole with a diameter of
  $2.65 \Lambda$.  (B) The field distribution of the fundamental mode
  in the airguiding photonic bandgap fibre, calculated at a normalized
  wavelength of $\lambda/\Lambda = 0.7$.}
\label{airguide}
\end{figure}

The question is how well a scalar calculation actually approximates
the results in photonic bandgap fibres, where the scalar wave equation
is obviously not a correct representation of the actual physical
problem. In order to examine this, we have chosen to apply our
finite-difference method to the extreme case of airguiding photonic
bandgap fibres. The dielectric structure used in our calculation is
shown in figure \ref{airguide}a. It has a width of $6 \sqrt{3}
\Lambda$ and consists of an air-silica cladding structure with a
holediameter of $D= 0.88 \Lambda$. The core defect is made by
inserting a large air hole with a diameter of $2.65 \Lambda$.  For the
calculation, we discretized the structure in $256 \times 256$ grid
points. We have chosen this specific structure, because it is known to
support guided modes in a full-vectorial calculation
\cite{broeng_airguide}.

A disadvantage of this finite-difference implementation is that
there is no simple way of calculating the position of the photonic
bandgaps. Therefore, we have used a full-vectorial planewave-method to
calculate the bandaps of an infinite triangular lattice with a hole
diameter of $0.88 \Lambda$. Once the position of the photonic bandgaps
have been located, it is possible to search for a defect mode. As
already mentioned, the \emph{EIGS}-command in \emph{Matlab} has the
useful ability to find eigenvectors around a specified
eigenvalue. This is particularly useful for bandgap fibres, since the
defect mode appears as an isolated eigenmode inside the boundaries of
the photonic bandgap.

By choosing a normalized wavelength of $\lambda/\Lambda = 0.7$ and
searching for an eigenvalue for which $\beta \approx k$, the scalar
finite-difference method finds a defect mode localized to the core
region. The field distribution of this defect mode is shown in figure
\ref{airguide}b. For simplicity we have chosen to depict the absolute
value of the field distribution. Therefore, the 6 lobes surrounding
the central defect do in fact have the opposite amplitude of the field
inside the defect. These 6 resonances are a common feature of airguiding
fibres, and they are also found in a full-vectorial calculation.

\begin{figure}[htpb]
\begin{center}
\epsfig{file=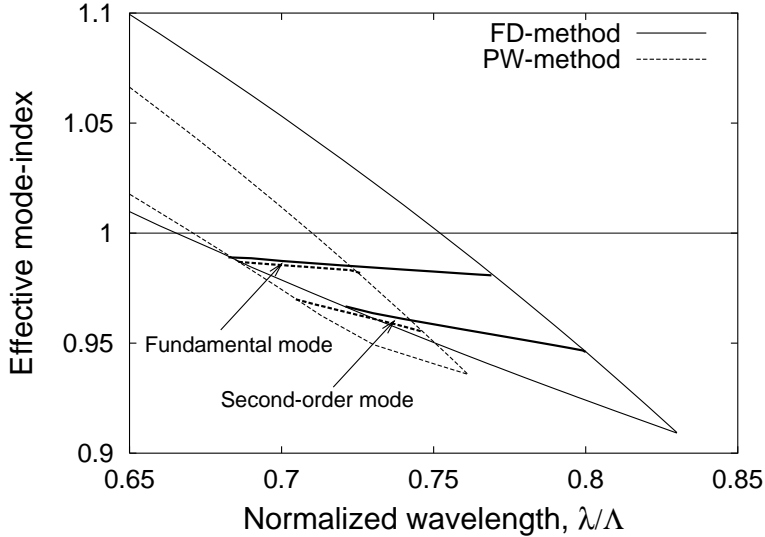, width=0.8\columnwidth,clip}
\end{center}
\caption{Comparison of the scalar finite-difference method (FD-method)
  and the full-vectorial planewave-method (PW-method) for an
  airguiding photonic bandgap fibre. A strong resemblance between the
  methods is found for both the fundamental and the second-order mode,
  but the finite-difference is seen predict reasonably wider
  bandgaps.}
\label{airguidecompare}
\end{figure}

In figure \ref{airguidecompare} we have mapped out the effective
mode-index for a range of the normalized wavelength. For comparison we
have included the guided modes and the bandgap edges from a
full-vectorial calculation. Both methods are found to predict the
existence of a fundamental and a second-order mode, and a reasonable
agreement is found between the results of the two methods. However, we
generally find that the finite-difference method overestimates the
values of the effective mode-index.

The major difference between the full-vectorial and the scalar
calculation is that the latter is seen to predict significantly
increased bandgaps. This naturally gives rise to a much wider range in
which the structure supports a confined mode. The bandedges in the
scalar calculation have been found as the modes that lie just above
and just below the defect modes. As the bandedge-modes are infact
cladding modes, and thus well distributed over the entire cross
section of the dielectric structure, they are more affected by the
truncation of the $\boldsymbol \Theta$-matrix than the well confined
defect modes. We have tried to increase the width of the calculation
domain and also the number of grid points, but this is not found to
have any influence on the overall result. It is therefore believed
that the increased bandgap size is a consequence of the scalar
approximation.

The final result indicates, that although a scalar approach provides
great insight to the basic physics of photonic bandgap fibres, it cannot
reveal the complete picture. This is not really surprising.  However,
we still believe that this method is of great interest, mainly due to
it simple implementation. Also, the model can easily be expanded to
include periodic boundary conditions and a full-vectorial implementation is
feasible as well.

\section{Conclusion}
The field of photonic crystal fibres has by now existed for almost a
decade and the results of the research efforts will probably soon move
inside the classroom and also be found in text-book material on fibre
optics and electro-magnetic theory of photonic crystals. This also
calls for simple approaches to modelling of micro-structured optical
fibres which are easy to implement and which without too much effort
can produce results which are in qualitative agreement with those
observed in real micro-structured fibres. We believe that the present
work provides such a simple approach.

In order to limit the complexity of the mathematical formulation of
the problem, we have considered a scalar wave-equation which is solved
by means of the most simple numerical approach to differential
equations; the finite-difference approximation. With appropriate
boundary conditions this results in a matrix eigenvalue problem. The
matrix, $\boldsymbol\Theta$, of this problem is highly sparse and has
very simple analytical matrix elements which only depend on the
lattice spacing, the frequency, and the dielectric function. Thus, no
implementation of complicated basis functions is required. For a given
frequency $\omega$ the propagation constant $\beta$ can easily be
found by finding the eigenvalues of $\boldsymbol\Theta$. By the aid of
standard numerical routines, we are able to solve for a specific
number of eigenvalues closest to a specified value. This is
particularly useful for bandgap guiding fibres, where the modes of
interest corresponds to either the smallest or the largest eigenvalue,
but are placed in an interval determined by the photonic bandgap edges.

In conclusion we find that the presented finite-difference method,
apart from being simple to implement and despite the extremely simple
model, is able to provide qualitative correct results for both
index-guiding and photonic bandgap guiding fibres. The latter case is
quite surprising, since modelling of photonic bandgap effects is
normally associated with full-vectorial methods. Furthermore, we find
that the method is robust and it is relatively simple to incorporate
periodic boundary conditions, or to expand the model to a
full-vectorial method. This holds interesting prospects for a future
development of this method.

\section*{Acknowledgements}

We acknowledge useful discussions with M.~D. Nielsen  and
B.~T. Kuhlmey. J.~L. is financially supported by
the Danish Technical Research Council (STVF).

\end{document}